\documentclass[prd,twocolumn,floats,floatfix,nofootinbib,superscriptaddress,showpacs]{revtex4}
\usepackage{graphicx}
\usepackage{graphics}
\usepackage{dcolumn}
\usepackage{amssymb}
\usepackage{bm}
\usepackage{amsmath}
\usepackage[dvips]{color}
\def\spose#1{\hbox to 0pt{#1\hss}}

\def\lta{\mathrel{\spose{\lower 3pt\hbox{$\mathchar"218$}}
     \raise 2.0pt\hbox{$\mathchar"13C$}}}
\def\gta{\mathrel{\spose{\lower 3pt\hbox{$\mathchar"218$}}
     \raise 2.0pt\hbox{$\mathchar"13E$}}}

\newcommand{\be}{\begin{equation}}
\newcommand{\en}{\end{equation}}
\newcommand{\bea}{\begin{eqnarray}}
\newcommand{\ena}{\end{eqnarray}}

\newcommand{\dee}{\mathrel{\mathop:}=}

\def\1{1 \!\! 1}

\def\ra{\rangle}
\def\la{\langle}
\newcommand{\ket}[1]{|{#1}\ra}

\newcommand{\bra}[1]{\la {#1}|}

\begin{document}
\title{Is there a super-selection rule in quantum cosmology?}

\author{E.~Sergio Santini}
\affiliation{CNEN - Comiss\~ao Nacional de Energia Nuclear \\
Rua General Severiano 90, Botafogo 22290-901,  Rio de Janeiro, Brazil}
\affiliation{CBPF - Centro Brasileiro de
Pesquisas F\'{\i}sicas, Rua Xavier Sigaud, 150, Urca,
CEP:22290-180, Rio de Janeiro, Brazil}

\date{\today}

\begin{abstract}
A certain approach to solving the Wheeler-DeWitt equation in quantum cosmology, which is based on a type of super-selection rule by which negative frequency solutions are discarded, is discussed. In a preliminary analysis, we recall well known results in relativistic quantum field theory, showing that adopt this approach of super-selection by discarding a sector of the frequencies, does not lead to acceptable results. In the area of quantum cosmology a qualitatively similar result is obtained: we show that by discarding solutions of negative frequencies, which is usually done in order to demonstrate ``strong'' results on the resolution of the singularity, important physical processes are lost, namely the existence of cyclic solutions which, under certain reasonable assumptions, can be interpreted as processes of creation-annihilation at the Planck scale, which are typical of any relativistic quantum field theory.
\end{abstract}


\maketitle

\section{Introduction}

The singularity problem in quantum cosmology has been addressed since the beginning of this area of research \cite{deWitt}. A long discussion on the possibility of resolve the singularity by mean of quantum effects took place since then.
In the context of the Wheeler-DeWitt (WDW) approach \cite{deWitt}\cite{halli3} , models have been obtained where the singularity persists in the quantum regime (see for example Ref. \cite{halli3}\cite{blyth}\cite{laflamme}\cite{hh1983}\cite{h1984}\cite{lemos}) and also models in which the singularity is avoided due to quantum effects (see Ref. \cite{kim}\cite{nelson1}\cite{nelson3}\cite{NSF}\cite{nelson4}\cite{kiefer}). That is to say there has been no absolute consensus on whether in the quantum regime the singularity is maintained as a strong result in the WDW approach. Furthermore there is no general agreement on the necessary criteria for quantum avoidance of singularities (for a clarifying analysis of the various criteria  see \cite{kiefer} and for the important problem of preservation of unitarity evolution see \cite{kim2}\cite{bertoni}). 

In the framework of Loop Quantum Cosmology (LQC) have been obtained a series of results where the big-bang singularity is avoided and substituted by a bounce coming from solutions of the equation of differences which is the equation that takes the place of the WDW equation (see Ref. \cite{bojowald}).\footnote{LQC has an important advantage over Wheeler-DeWitt quantum cosmology because has its foundations in the theory of Loop Quantum Gravity, where there are less conceptual problems than in the case of the canonical quantization that lead to the WDW equation.}
At the same time it has been claimed that the Wheeler-DeWitt approach to  quantum cosmology does not solve the singularity problem of classical cosmology (see for example \cite{CS}\cite{ash4}\cite{ash2}\cite{ash3}\cite{ash1}). This general assertion was already widely criticized in \cite{nosso}. There it was shown firstly that the assertion is not  precise because to address this question it is necessary to specify not only the quantum interpretation adopted but also the quantization scheme chosen. In second place it was demonstrated  that quantum bounces occur when one consider the Bohm-de Broglie interpretation in any of the two different usual quantization schemes: the Schr\"odinger-like quantization, which essentially takes the square-root of the resulting Klein-Gordon equation through the restriction to positive frequencies and their associated Newton-Wigner states, or the induced Klein-Gordon quantization, that allows both positive and negative frequencies together. We refer the  reader, interested  in this study, to the last cited reference.
The aim of this letter is to analyze  and to discuss the quantization scheme which involves the restriction to a single sector of frequencies (say discarding  the negative frequency solutions\footnote{The problem of negative frequencies in quantum cosmology is known since early works in the subject, see for example \cite{blyth}.})  which is made invoking a type of super-selection rule \cite{ash1}.
As a preliminary study we analyze what happens in a basic quantum field theory (QFT), i.e. a quantum free scalar field, when we discard the negative frequencies solutions. As it is well known the fundamental Lorentz symmetry is lost, or in other words we are lead to a violation of causality. 
We then study the WDW approach to quantum cosmology. At the beginning  we  briefly outline the arguments of \cite{ash1} for a super-selection rule, making some discussion and after that we develop the central part of this letter: we analyze the behavior of the bohmian trajectories obtained from the solutions of the WDW equation for a Friedmann-Lem\^aitre-Robertson-Walker (FLRW) model with flat spatial sections, assuming the content of matter of the universe as given by a free, massless, minimally coupled scalar
field, while negative frequencies are incorporated to the  positive frequency initial solution. This is implemented using a superposition of solutions modulated by two gaussian symmetrically located around $k=0$. The gaussian width is varied from an initial value representing an almost non-overlapped configuration (the gaussians are completely disjoint), which indicates a positive frequency solution, then going  through several increasing values representing a greater partial overlap, indicating a greater weight  of  negative frequencies in the integral, until we overcome a certain ``threshold'' (see below)  from which,  phenomena qualitatively different begins to occurs.
We were able to show that when the negative frequency solutions are incorporated beyond a certain value, a fundamental phenomena appear:  the existence of cyclic universe solutions which could be interpreted as  processes of creation-destruction of universes at Planck scale, provided the scalar field is assumed to play the role of a time. These phenomena  are usual in any QFT and we see that when the negative frequency solutions are not considered in quantum cosmology, theses processes are lost or, in other words, no cyclic universe solutions is present.

This letter is organized as follows: in section II as a preliminary study we analyze the case of a QFT, then in section III the case of quantum cosmology, in section IV the BdB quantum cosmology is studied and our results are obtained, in section V we present our conclusions.

\section{Discarding negative frequencies solutions in Klein-Gordon}
This section contains well known results in quantum field theory. But we wish to clarify the problem to be studied in the next section, using a model already known, which is a scalar field satisfying the Klein-Gordon equation and recreating the type of problem that can occur when the frequencies of a sector (say negative) are discarded from the general solution. The idea is to confront qualitatively with the model of quantum cosmology discussed in the next section. The model of a massless scalar field,  which would be most appropriate to compare with the quantum cosmological model of the next section, is subjected to the same analysis and satisfies the same results recreated here, since it can be obtained without problems by taking the limit of zero mass from the massive field considered here (always with zero spin) \cite{weinberg}Chap.5.9.

We consider a free scalar field $\psi$ satisfying the usual Klein-Gordon equation:

\be
\frac{\partial^2 \psi}{\partial t^2}-\nabla^2\psi + m^2\psi=0 \,\,\, .
\en

A general solution can be written as a sum of two terms:

\be \label{klein}
\psi= \int_{-\infty}^{\infty} dp \tilde{\psi}_{+}(p) e^{\frac{i}{\hbar}(px-Et)}+\int_{-\infty}^{\infty} dp \tilde{\psi}_{-} (k) e^{\frac{i}{\hbar}(px+Et)}
\en

being the first term the ``positive frequency solution" and the second one the ``negative frequency solution''.

As we know the energy $E$  satisfy (for a given momentum p)  the condition 
$ E^2=p^2 + m^2$, i.e. it can have two values, $\pm\sqrt(p^2 + m^2)$. In principle only positive values of $E$ can 
have the physical significance of the energy of a free particle. But the negative 
values cannot be simply omitted: the general solution of the wave equation can be 
obtained only by superposing all its independent particular solutions, being the negative solutions re-interpreted in the formalism of second quantization. 
We have a complete set of commuting observables given by the energy and the momentum\footnote{Indeed we must also know the helicity, which is zero, but we can ignore it without affecting our analysis.}:

\be
\hat{E}\equiv i\hbar\frac{\partial}{\partial t}
\en

\be
\hat{p}\equiv -i\hbar\frac{\partial}{\partial x}
\en
As it is known, these are ``even'' operators which means that they transform positive  frequency solutions in positive frequency solutions and negative  frequency solutions in negative 
frequency solutions. This feature does not allow us, in any way, to dispense one of the sectors (say negative). There is no selection rule that allows us to dispense with one of the sectors. If we did that, in first place  there will be no room for antiparticles, which means the absence of the rich processes of creation and annihilation. In second place it would not be possible to satisfy completely the Lorentz symmetry, 
more precisely, it will be violated the four-dimensional inversion (which is a four-rotation with determinant $=+1$), since its fulfillment  make necessary the simultaneous presence 
in the Eq. (\ref{klein}) of terms having both signs of $E$ in the 
exponents: these signs are changed by the substitution $t\rightarrow - t$(see \cite{LandauIV} section 11.)
With this,  a violation of CPT invariance would be arbitrarily introduced (see \cite{LandauIV} section 13. )

Another way to see the problem  is to analyze the  process consisting in a particle propagating from space-time point $x$ to space-time point $y$. The field $\psi^{*}(x)$ 
creates a particle at $x$, and $\psi(y)$ destroys a particle at $y$. The amplitude for this process is given by

\be \label{amp}
\bra{0}\psi(y)\psi^{*}(x)\ket{0}
\en
 and, at space-like separations (i.e. out of the light cone), it must vanish because no signal can propagate faster than light. 
Now, as we know  $\psi(y)\ket{0}=0$, because  the vacuum is annihilated by $\psi$,  then if the commutator $[\psi(y),\psi^{*}(x)]$  vanishes for 
space-like separated regions, i.e

\be \label{conm}
[\psi(y),\psi^{*}(x)]=0  \, \, \, \, 
for \, \,
 (x-y)^2 < 0 \, \, ,
\en
 then the amplitude (\ref{amp}) will indeed vanish at space-like separations. In other words,  (\ref{conm}) is a sufficient condition for the amplitude (\ref{amp}) vanish at space-like separations.
In computing the commutator in (\ref{conm}) we can use the plane wave expansion for the field operators $\psi$ and $\psi^{*}$. If we assume that this expansion involves   a sum over plane waves with only positive frequencies (as in the case of  non relativistic free fields) then it is not mathematically possible to adjust the coefficients of those expansions in such a way that they verify (\ref{conm}), unless they commute identically for all the space-time. It is  necessary to allow negative frequency plane waves in the field expansions in order to satisfy (\ref{conm}), i.e commutation  at space-like separations but not everywhere. By discharging the negative frequencies  leads to a violation of causality, \cite{weinberg}\cite{hatfield}.

\section{Discarding negative frequencies solutions in quantum cosmology}
We pointed out in the introduction that in several papers the approach  WDW to quantum cosmology has been criticized, because they show that it is not possible solve the big bang singularity. We have already noted that this strong statement has been criticized in \cite{nosso}. But there is an argument of ``super-selection rule'' used to arrive at that statement, we want to discuss now. We are going to analyse the validity of the procedure, which involves working with a single sector of frequencies.
Here we outline  the argument of reference \cite{ash1}: In that reference  is studied the WDW limit of Loop Quantum Cosmology (LQC) by working at the regime where effects of quantum discrete geometry can be neglected. The WDW equation obtained has the same form as the Klein-Gordon equation in a static space-time: 

\be
\frac{\partial
^2\Psi}{\partial \phi ^2}  +\Theta \Psi=0 \quad,
\en

where the field $\phi$ plays the role of time and $\Theta$ of the spatial laplacian given by Eq. (3.4) of \cite{ash1}:

\be
 \Theta \equiv -\frac{16\pi G}{3B(\mu)} \frac{\partial}{\partial \mu}\sqrt{\mu}\frac{\partial }{\partial \mu} 
\en

where $\mu$ is the spatial coordinate and $B(\mu)$ is an eigenvalue of the operator $\widehat{|\mu|^{-3/2}}$, within a multiplicative constant. A general solution is obtained as a superposition of positive and negative frequencies:

\bea
\Psi(\mu, \phi)&=&\int_{-\infty}^{+\infty}dk \tilde{\Psi}_{+}e_{k}(\mu)e^{i\omega \phi} + \int_{-\infty}^{+\infty}dk \tilde{\Psi}_{-}\bar{e}_{k}(\mu)e^{-i\omega \phi}\nonumber\\
&=& \Psi_{+}(\mu, \phi) +\Psi_{-}(\mu, \phi) \quad.
\ena

where $e_{k}(\mu)$ are the eigenvectors of $\Theta$ with eigenvalues $\omega^2$. A complete set of Dirac Observables is given by

\be
\hat{p}_{\phi}\equiv-i\hbar \frac{\partial \Psi}{\partial \phi}
\en

\bea
\widehat{|\mu|_{\phi_{0}}}\Psi(\mu,\phi)\equiv e^{i\sqrt{\Theta}(\phi-\phi_{0})}|\mu|\Psi_{+}(\mu,\phi_{0}) + \nonumber\\e^{-i\sqrt{\Theta}(\phi-\phi_{0})}|\mu|\Psi_{-}(\mu,\phi_{0}) \quad,
\ena

and we quote verbatim from \cite{ash1}:``
both these operators preserve the positive and negative frequency subspaces. Since they constitute a complete family of Dirac observables, we have {\it{superselection}}. In quantum theory we can restrict ourselves to one superselected sector. We focus on the positive frequency sector, and from now on, drop the suffix $+$.''

The restriction to positive frequencies sector through a type of ``super-selection rule'' allows build a Hilbert space ``physical'' which is the space of wave functions of positive frequency with finite norm (the norm given by equation (3.15) of \cite{ash1}).
This may be correct from a mathematical point of view, especially because it allows to show that the evolution WDW does not resolve the singularity (section III B of \cite{ash1}).
However, in addition to the fact that the issue of the WDW singularity resolution, as presented in \cite{ash1}, seems to depend on the inclusion
or not inclusion of negative frequencies, it does not seem like a good way to follow as this procedure makes  loss of important physics
characteristics associated with the existence of negative
frequency solutions, as we are going to show in section IV.

\section{The Bohm-De Broglie theory applied to quantum cosmology.}\label{dBBqc}

The Bohm-De Broglie quantum theory (see \cite{dBB}) can be consistently implemented in quantum cosmology (see \cite{kowalski2}).
Considering homogeneous mini-superspace models, which have a finite number
of degrees of freedom, the general form of the associated Wheeler-De Witt equation reads
\begin{equation}
\label{bsc}
-\frac{1}{2}f_{\rho\sigma}(q_{\mu})\frac{\partial \Psi (q)}{\partial q_{\rho}\partial q_{\sigma}}
+ U(q_{\mu})\Psi (q) = 0 \quad,
\end{equation}
where $f_{\rho\sigma}(q_{\mu})$ is the minisuperspace DeWitt metric of the model, whose inverse is denoted by $f^{\rho\sigma}(q_{\mu})$. By writing the wave function in its polar form, $\Psi = R \ e^{iS}$, the complex equation (\ref{bsc}) decouples in two real equations
\begin{equation}
\label{hoqg}
\frac{1}{2}f_{\rho\sigma}(q_{\mu})\frac{\partial S}{\partial q_{\rho}}
\frac{\partial S}{\partial q_{\sigma}}+ U(q_{\mu}) + Q(q_{\mu}) = 0 \quad,
\end{equation}
\begin{equation}
\label{hoqg2}
f_{\rho\sigma}(q_{\mu})\frac{\partial}{\partial q_{\rho}}
\biggl(R^2\frac{\partial S}{\partial q_{\sigma}}\biggr) = 0 \quad,
\end{equation}
where
\begin{equation}
\label{hqgqp}
Q(q_{\mu}) \dee -\frac{1}{2R} f_{\rho\sigma}\frac{\partial ^2 R}
{\partial q_{\rho} \partial q_{\sigma}}
\end{equation}
is called the quantum potential. The Bohm-De Broglie interpretation applied to quantum cosmology states that the trajectories $q_{\mu}(t)$ are real, independently of any observations. Equation (\ref{hoqg}) represents their Hamilton-Jacobi equation, which is the classical one  added with a quantum potential term Eq.(\ref{hqgqp}) responsible for the quantum effects. This suggests to define

\begin{equation}
\label{h}
\pi^{\rho} = \frac{\partial S}{\partial q_{\rho}} ,
\end{equation}
where the momenta are related to the velocities in the usual way

\begin{equation}
\label{h2}
\pi^{\rho} = f^{\rho\sigma}\frac{1}{N}\frac{\partial q_{\sigma}}{\partial t} ,
\end{equation}

being $N$ the lapse function. In order to obtain the quantum trajectories, we have to solve the following
system of first order differential equations, called the guidance relations
\begin{equation}
\label{h3}
\frac{\partial S(q_{\rho})}{\partial q_{\rho}} =
f^{\rho\sigma}\frac{1}{N}\dot{q}_{\sigma} .
\end{equation}

The above equations (\ref{h3}) are invariant under time re-parametrization. Therefore, even at the quantum level, different time gauge choices of $N(t)$ yield the same space-time geometry for a given non-classical solution $q_{\alpha}(t)$. Indeed, there is no problem of time in the de~Broglie-Bohm interpretation for
minisuperspace quantum cosmological models \cite{bola27}. However, this is no longer true when one considers the full superspace (see \cite{santini0}\cite{tese}). Notwithstanding, even with the problem of time in the superspace, the theory can be consistently formulated (see \cite{tese}\cite{cons}).

Let us then apply this interpretation to our minisuperspace model, which is given by a spatially flat Friedmann (FLRW) universe with a massless free scalar field.
The Wheeler-DeWitt equation reads\footnote{This is the same model studied in \cite{nosso} Sec.II and IV}

\begin{equation}
\label{wdw}
-\frac{\partial ^2\Psi}{\partial \alpha ^2} +  \frac{\partial
^2\Psi}{\partial \phi ^2} = 0 \quad,
\end{equation}

where $\phi$ is the scalar field and  $\alpha \equiv \log{a}$, being $a$ the scale factor. Comparing Eq.~(\ref{wdw}) with Eq.~(\ref{bsc}), we obtain,
from Eqs.~(\ref{hoqg}) and (\ref{hoqg2}),

\begin{equation}
\label{hoqgp}
- \biggl(\frac{\partial S}{\partial \alpha}\biggr)^2 + \biggl(\frac{\partial S}{\partial \phi}\biggr)^2
+ Q(q_{\mu}) = 0 \quad,
\end{equation}
\begin{equation}
\label{hoqg2p}
\frac{\partial}{\partial \phi} \biggl(R^2\frac{\partial S}{\partial \phi}\biggr)
- \frac{\partial}{\partial \alpha} \biggl(R^2\frac{\partial S}{\partial \alpha}\biggr) = 0 \quad,
\end{equation}
where the quantum potential reads
\begin{equation}
\label{qp1}
Q(\alpha ,\phi )\dee \frac{1}{R}\biggr[\frac{\partial^{2}R}
{\partial \alpha^{2}}-\frac{\partial^{2}R}{\partial \phi^{2}}\biggl]\quad .
\end{equation}
The guidance relations (\ref{h3}) are
\begin{equation}
\label{guialpha}
\frac{\partial S}{\partial \alpha}=-\frac{e^{3\alpha}\dot{\alpha}}{N}\quad ,
\end{equation}
\begin{equation}
\label{guiphi}
\frac{\partial S}{\partial \phi}=\frac{e^{3\alpha}\dot{\phi}}{N}\quad .
\end{equation}

We can write equation Eq.~(\ref{wdw}) in null coordinates,
\begin{eqnarray}
\label{nulas}
v_l\dee\frac{1}{\sqrt{2}}(\alpha+\phi) \quad & &\quad  \alpha\dee\frac{1}{\sqrt{2}} \left(v_l+v_r\right)\nonumber\\
v_r\dee\frac{1}{\sqrt{2}}(\alpha-\phi) \quad  & &\quad  \phi\dee\frac{1}{\sqrt{2}} \left(v_l-v_r\right)
\end{eqnarray}
yielding,
\begin{equation}
\left(-\frac{\partial^{2} }{\partial v_l \partial v_r }\right) \Psi \left(v_l,v_r \right) =0 \quad.
\end{equation}
The general solution is
\begin{equation}
\label{sol0}
\Psi(u,v) = F(v_l) + G(v_r) \quad,
\end{equation}
where $F$ and $G$ are arbitrary functions.
Using a separation of variable method, one can write these solutions
as Fourier transforms given by

\begin{equation}
\Psi(v_l,v_r) = \int_{-\infty}^{\infty} d k U(k)\ e^{ikv_l} + \int_{-\infty}^{\infty} d k V(k)\ e^{ikv_r} \quad,
\end{equation}
being $U$ and  $V$ also  two arbitrary functions,
or, because for our purpose is better to work in the original coordinates $\alpha$ and $\phi$:

\begin{equation}
\label{sol0k}
\Psi(\alpha,\phi) = \int_{-\infty}^{\infty} d k U(k)\ e^{ik\frac{1}{\sqrt{2}}(\alpha+\phi)} + \int_{-\infty}^{\infty} d k V(k)\ e^{ik\frac{1}{\sqrt{2}}(\alpha-\phi)} \quad.
\end{equation}

For our numerical analysis we take the arbitrary functions $U(k)$ and $V(k)$ as being the gaussians

\be
U(k)=e^{-(k-d)^2/\sigma^2}
\en
\be
V(k)=e^{-(k+d)^2/\sigma^2}
\en

then we have

\bea
\label{sol0kG}
\Psi(\alpha,\phi) = \int_{-\infty}^{\infty} d k  e^{-(k-d)^2/\sigma^2}\ e^{ik\frac{1}{\sqrt{2}}(\alpha+\phi)} + \nonumber\\ \int_{-\infty}^{\infty} d k  e^{-(k+d)^2/\sigma^2}\ e^{ik\frac{1}{\sqrt{2}}(\alpha-\phi)} \quad.
\ena

After integration and within a normalization\footnote{In our study, the normalization of the wave function is irrelevant  because we are going to extract information only from its phase.} constant, we have:

\bea
\Psi(\alpha,\phi)=&&\nonumber\\|\sigma|\sqrt{\pi} e^{i\frac{d\phi}{\sqrt{2}}-\frac{\sigma^2(\alpha^2+\phi^2)}{8}} \left\{e^{\frac{id\alpha}{\sqrt{2}}-\frac{\sigma^2\alpha\phi}{4}} + e^{-\frac{id\alpha}{\sqrt{2}}+\frac{\sigma^2\alpha\phi}{4}}  \right\} \,\,.
\ena

To obtain the quantum trajectories it is necessary to calculate the phase $S$ of the above  wave function and substitute it into the guidance equations. We will work in the gauge $N = 1$. Computing the phase we have:

\be
S=\frac{d\phi}{\sqrt{2}}+ \arctan(\tanh(\frac{\sigma^2\alpha\phi}{4}) \tan(\frac{d\alpha}{\sqrt{2}})) \quad,
\en

which, after substitution in (\ref{guialpha}) and (\ref{guiphi}), yields a planar system given by:

\begin{equation}\label{guia1}
\dot{\alpha}=\frac{\phi \sigma^2 \sin(\sqrt{2}d\alpha) +2\sqrt{2}d \sinh(\frac{\sigma^2\alpha\phi}{2})}
{e^{3\alpha}4[\cos(\sqrt{2}d\alpha)+\cosh(\frac{\sigma^2\alpha\phi}{2})]}
\end{equation}

\begin{equation}\label{guia2}
\dot{\phi}=\frac{2\sqrt{2}d \cosh(\frac{\sigma^2\alpha\phi}{2})+2\sqrt{2}d \cos(\sqrt{2}d\alpha)-\alpha\sigma^2 \sin(\sqrt{2}d\alpha)}
{e^{3\alpha}4[\cos(\sqrt{2}d\alpha)+\cosh(\frac{\sigma^2\alpha\phi}{2})]} \quad.
\end{equation}

\begin{figure}[!t]
\centering
\includegraphics[height=80mm,width=80mm,angle=270]{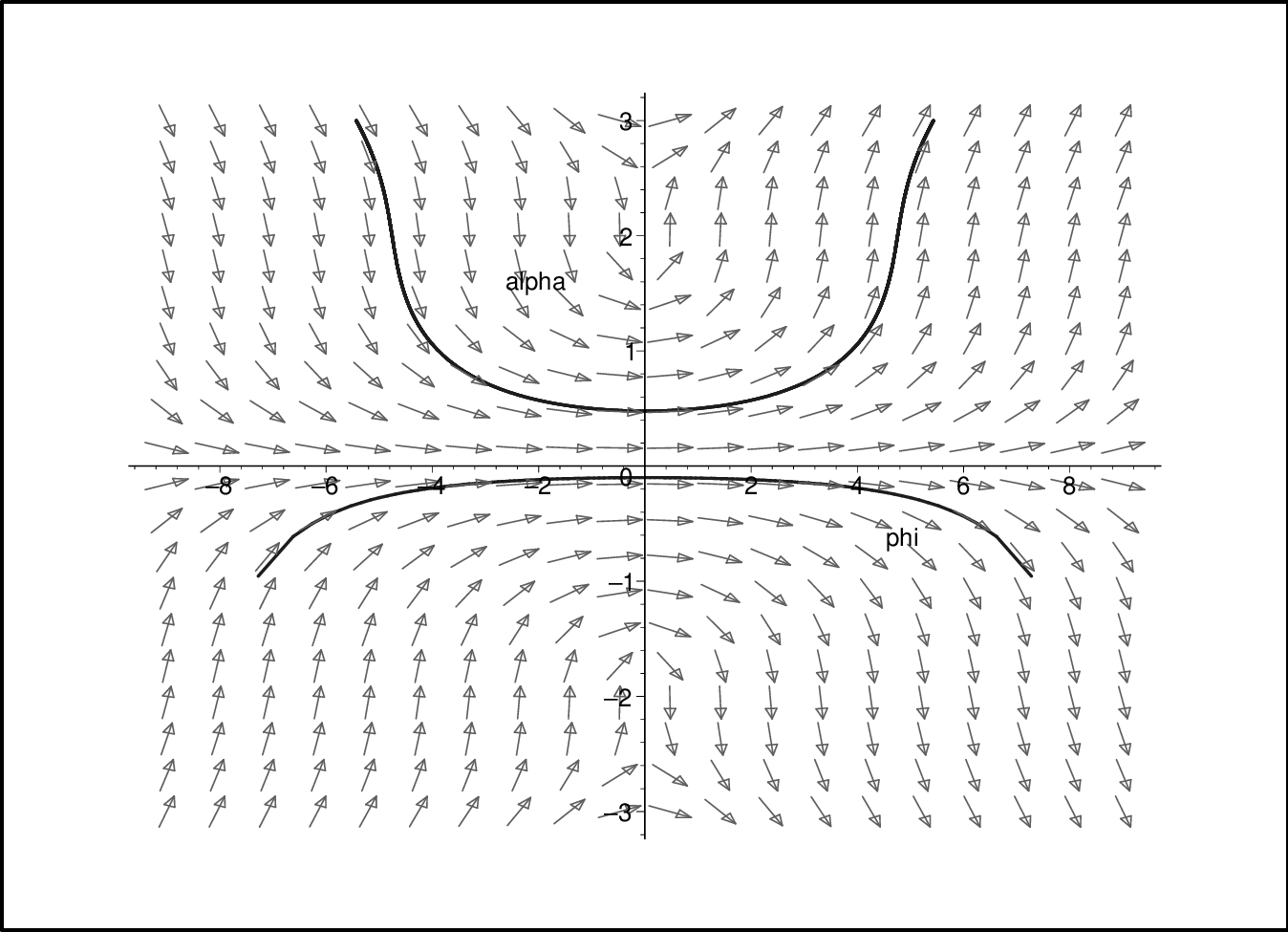}
\caption{The field plot shows the family of trajectories for Bohm guidance  equations (\ref{guia1}),(\ref{guia2}) associated to a wave functional with positive frequencies only. Two of them that describe their general behavior are depicted in solid line: the first representing a bouncing universe while the second one corresponds to a universe which begins and ends in singular states (``big bang - big crunch'' universe). }
\label{traj}
\end{figure}

Equations (\ref{guia1}),(\ref{guia2}) give the direction of the geometrical tangents
to the trajectories which solves this planar system.
By plotting the tangent direction field, it is possible to obtain the trajectories (Fig. \ref{traj}).
The line $\alpha=0$ divides the configurations space in two symmetric regions and the line $\phi=0$ contains all the singular points, which are  nodes and centers.  The nodes arise when the denominator in the above equations, proportional to the norm of the wave function, is zero. No trajectory pass through these points. They happen when $\phi=0$ and $\alpha= (2n+1)\frac{\pi}{\sqrt{2}d}$, $n$ an integer, with periodicity $\frac{\sqrt{2}\pi}{d}$. The center points appear when the numerators are zero.  They are given by
$\phi=0$ and
$\alpha = \frac{2\sqrt{2}d}{\sigma^2}\cot(\frac{\sqrt{2}}{2}d \alpha)$.

\vspace{1cm}

In Ref. \cite{nosso}, the bohmian trajectories, corresponding to solutions of equation WDW of positive frequency, were obtained, such as shown in figure \ref{traj}. Here is possible to distinguish two kind of trajectories. The upper half of the figure contains trajectories describing bouncing universes while the lower half corresponds to universes that begins and ends in singular states (``big bang - big crunch'' universe).

\subsection{Positive and negative frequencies}
If we allows negative frequencies and positive frequencies in the solution, it is possible to observe the occurrence of cyclic universes which are given by oscillatory trajectories in $\phi$, as we will see. In this case, if one wishes to interpret $\phi$ as time, this corresponds to creation and annihilation of expanding and contracting universes that exist for a very short duration\footnote{At this point we must to note that a continuity equation for the ensemble of trajectories with an certain  distribution function of initial conditions, is absent. For a discussion of this point, see \cite{nosso}, section IV.}.

In order to get an idea of how the inclusion of negative frequencies works in the behavior of solutions, we  gradually introduce them. 

\begin{figure}[ht]
\includegraphics[height=80mm,width=80mm,angle=270]{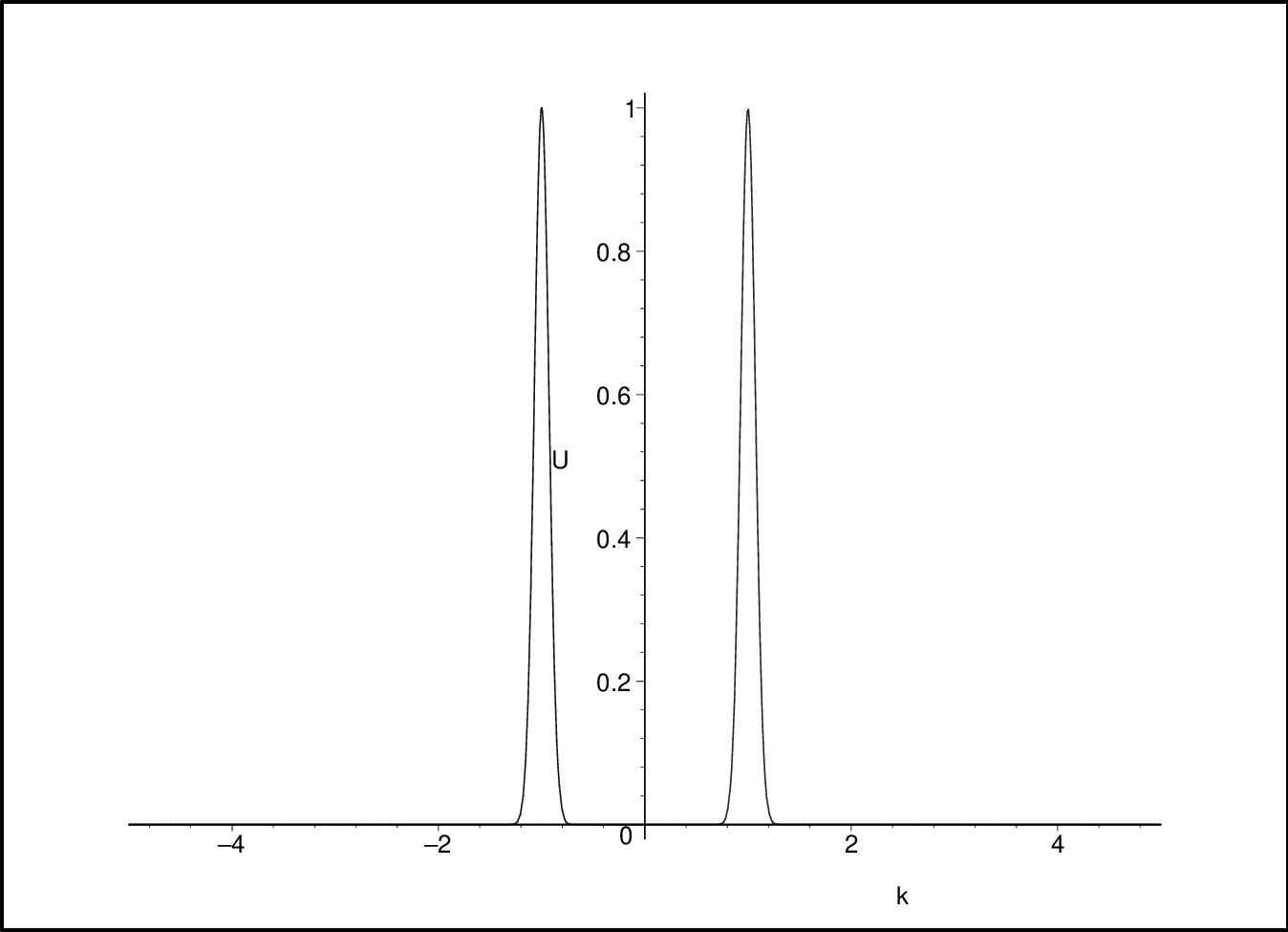}
\caption{$U(k)$ and $V(k)$ are two sharply peaked gaussians centered at $k=d$ and $k =-d$ respectively. This give a positive frequency solution, equation (\ref{solf+2}).}
\label{gd}
\end{figure}

We start choosing the values $\sigma$ and $d$ in (\ref{sol0kG}) with  $\sigma << 1$ and $d \geq 1$. In this case the functions   $U(k)$ and $V(k)$ are two sharply peaked gaussians centered at $k=d$ and $k =-d$ respectively (figure \ref{gd}). Then the wave function (\ref{sol0kG})  can be written very approximately as

\bea
\label{solf+2}
\Psi(\alpha,\phi) \approx \int_{0}^{\infty} d k  e^{-(k-d)^2/\sigma^2}\ e^{ik\frac{1}{\sqrt{2}}(\alpha+\phi)} + \nonumber\\ \int_{-\infty}^{0} d k  e^{-(k+d)^2/\sigma^2}\ e^{ik\frac{1}{\sqrt{2}}(\alpha-\phi)} \quad,
\ena

which means a positive frequency solution (note that $\phi$, i.e. the ``time'', appears in the exponential only with a positive sign).

\begin{figure}[ht]
\includegraphics[height=80mm,width=80mm,angle=270]{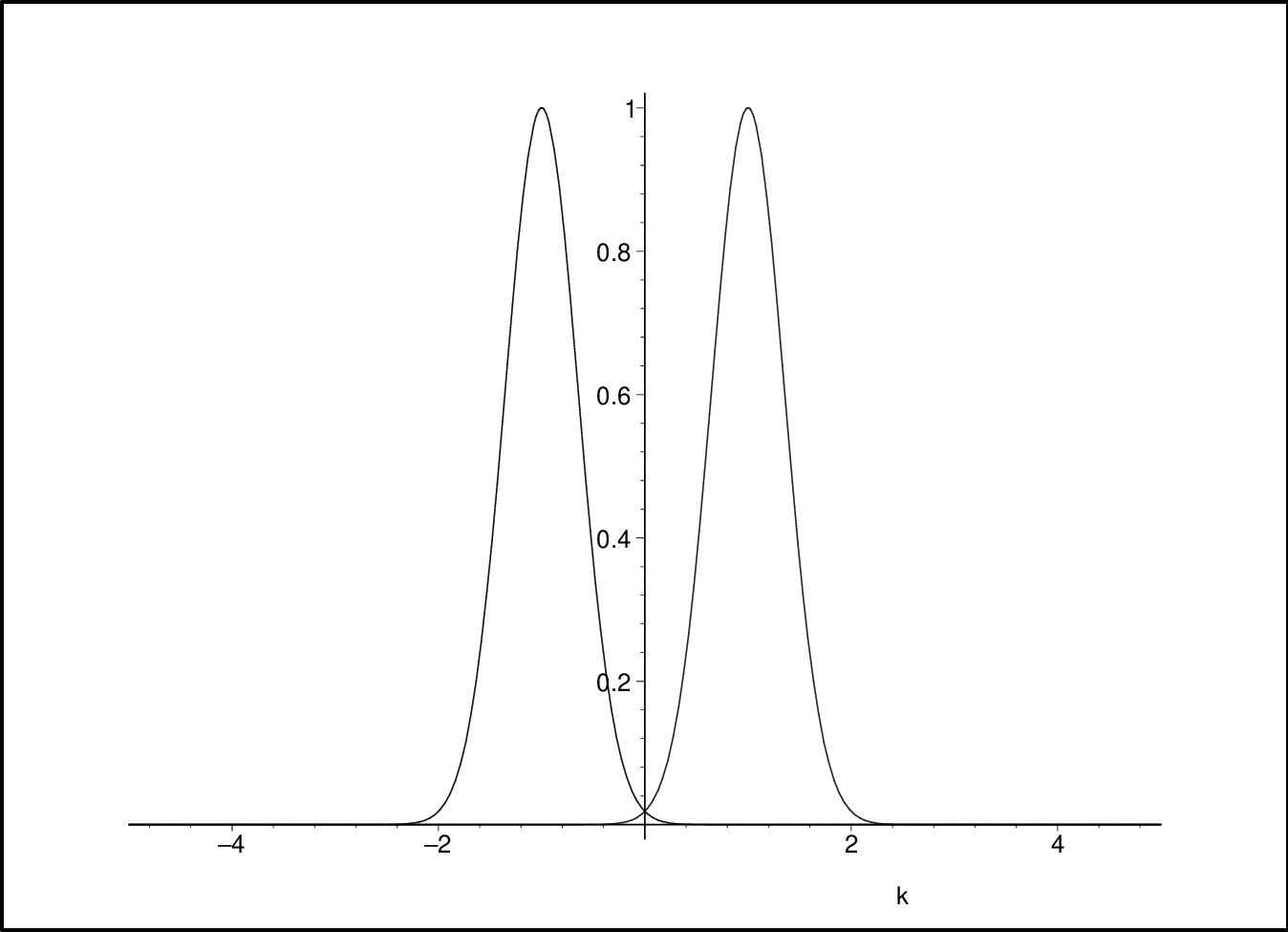}
\caption{$U(k)$ and $V(k)$ can no longer   be considered almost disjoint but will begin to overlap. Negative frequencies will begin to have an appreciable  weight in the integral.}
\label{go}
\end{figure}

If we increase sufficiently the parameter $\sigma$ (the ``width'')  the two gaussians  can no longer   be considered almost disjoint but will begin to overlap (figure \ref{go}). This means that the approximation (\ref{solf+2}) is no longer valid and negative frequencies will begin to be included with a greater weight in the integral. As we continue to increase the parameter $\sigma$ more and more negative frequencies will have a greater weight  in the solution.

\begin{figure}[!ht]
\includegraphics[height=80mm,width=80mm,angle=270]{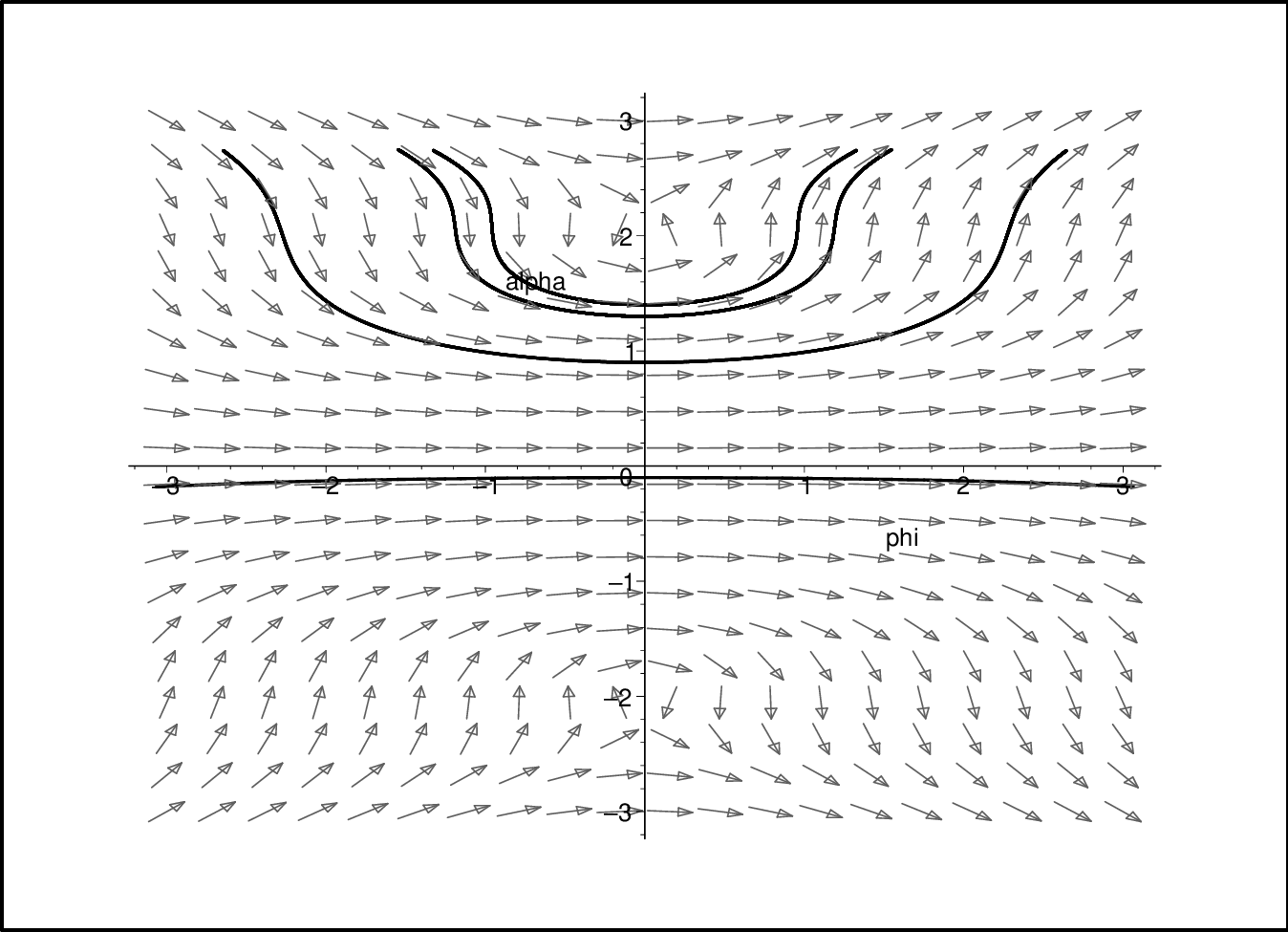}
\caption{The field plot shows the family of trajectories for the Bohm guidance equations (\ref{guia1}),(\ref{guia2}) associated to a wave functional with only the positive frequencies solutions. $\sigma=0.5$ and $d=1$. }
\label{3}
\end{figure}

\begin{figure}[!ht]
\includegraphics[height=80mm,width=80mm,angle=270]{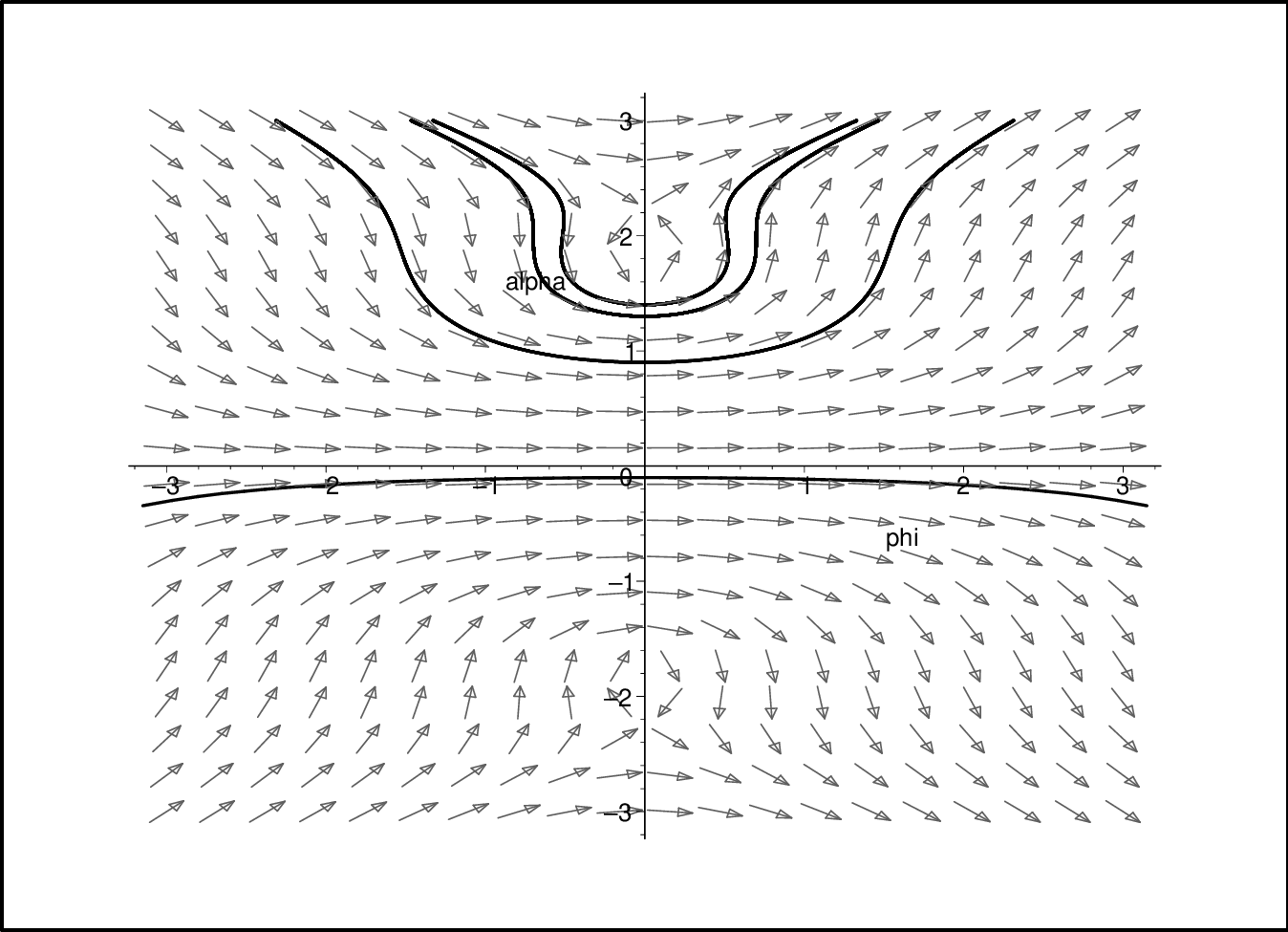}
\caption{The field plot shows the family of trajectories for the Bohm guidance equations (\ref{guia1}),(\ref{guia2}) associated to a wave functional with  the positive frequencies and a bit of negative frequencies, which begin to weigh on the integral . $\sigma=0.7$ and $d=1$. }
\label{4}
\end{figure}

\begin{figure}[!ht]
\includegraphics[height=80mm,width=80mm,angle=270]{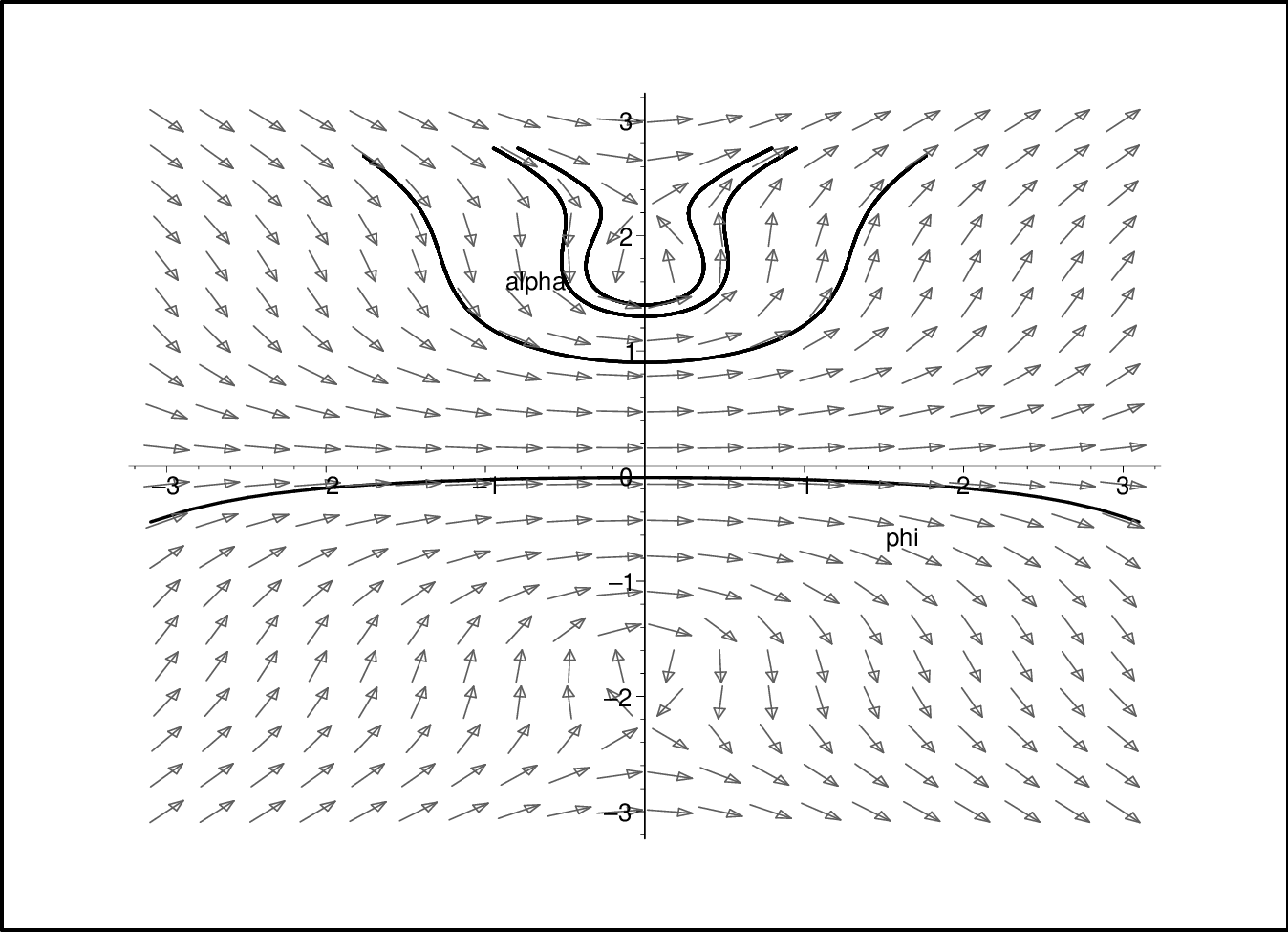}
\caption{The field plot shows the family of trajectories for the Bohm guidance equations (\ref{guia1}),(\ref{guia2}) associated to a wave functional with  the positive frequencies and more an more negative frequencies which begin to weigh on the integral. $\sigma=0.8$ and $d=1$.}
\label{5}
\end{figure}

We solved the Bohm guidance equations (\ref{guia1}),(\ref{guia2}) and obtained the bohmian trajectories for several increasing values or the parameter $\sigma$ (parameter $d$ was kept constant) and for the same initial conditions. The results are depicted in figures from \ref{3} to \ref{7}.

In figure \ref{3} we have practically only positive frequencies solutions, in figure \ref{4} negative frequencies solutions begin to weigh on the integral, in figure \ref{5}  a little more negative frequencies weigh in the integral . We're adding more and more negative frequencies (this is because each gaussian has an increasingly significant tail on the semi-axis which is opposite to the one containing its center) until we see that at a certain point, a trajectory becomes cyclical (Fig. \ref{6}). Here there seems to be a threshold for this particular trajectory, which will be discussed below. At the end  in figure \ref{7} positive and negative frequencies appear, in some sense,  alike, as would be in a general solution of the  WDW equation. In this last figure we observe the occurrence of cyclic universes where, as we said, we can interpret as processes of the creation and destruction of universes, if we accept $\phi$ fulfilling the role of a time. This situation of creation and annihilation of universes is a typical
feature of a relativistic quantum field theory. After all, Wheeler-DeWitt equation already represent a second-quantized field theory. As such, it is expected that creation-annihilation processes occur naturally. This fundamental processes are lost along the demonstrations using  the ``super-selection rule'', then, in our humble understanding, we think that ``rule'' does not exist.

\subsection{A threshold for the emergence of cyclical universes?}

\begin{figure}[!ht]
\includegraphics[height=80mm,width=80mm,angle=270]{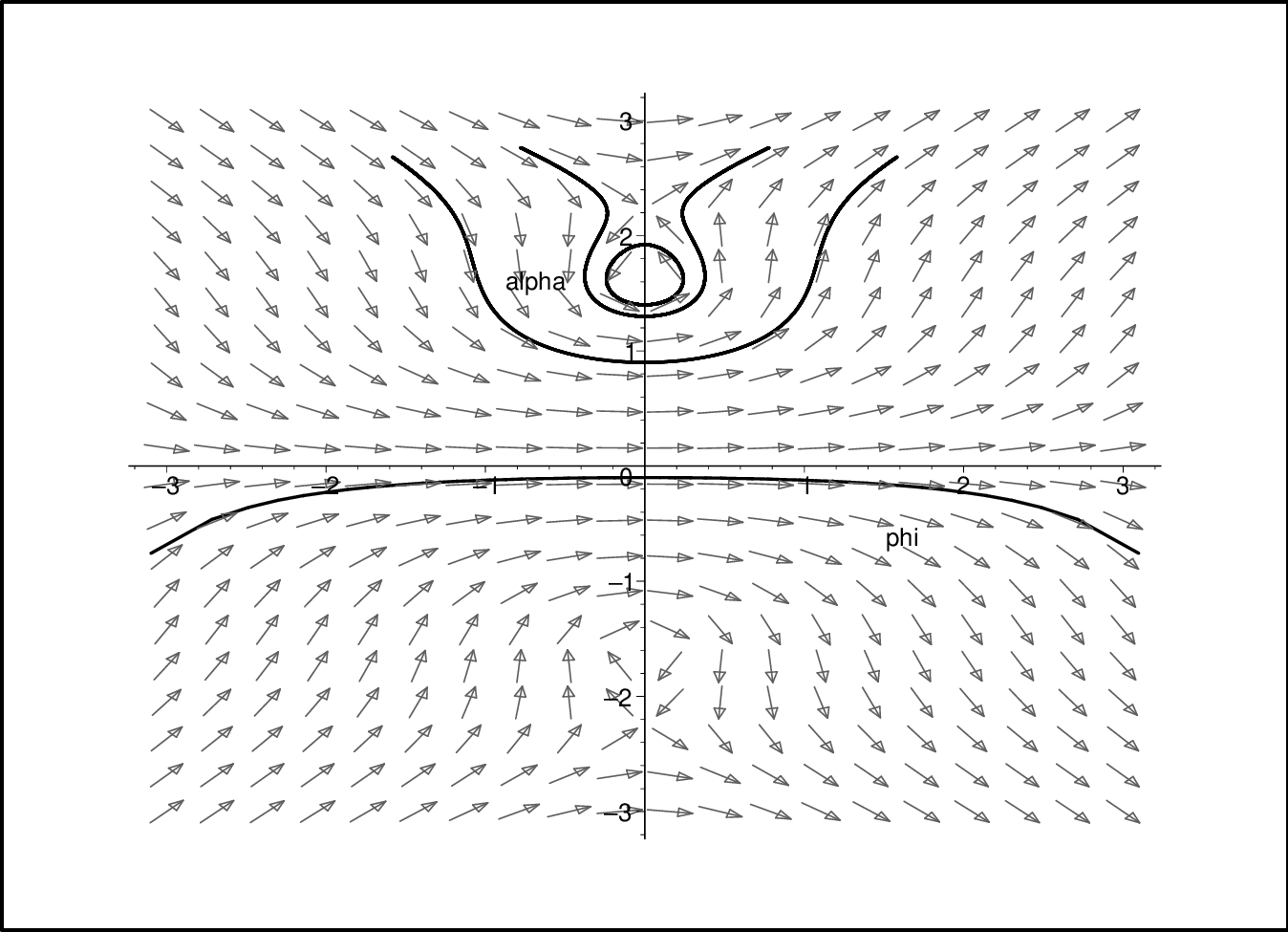}
\caption{The field plot shows the family of trajectories for the Bohm guidance equations (\ref{guia1}),(\ref{guia2}) associated to a wave functional with the positive frequencies and with such weigh of the negative frequencies that a cyclic universe is formed (trajectory passing through $\alpha=1.4 $, $\phi=0 $ ). $\sigma=0.9$ and $d=1$. }
\label{6}
\end{figure}

\begin{figure}[!ht]
\includegraphics[height=80mm,width=80mm,angle=270]{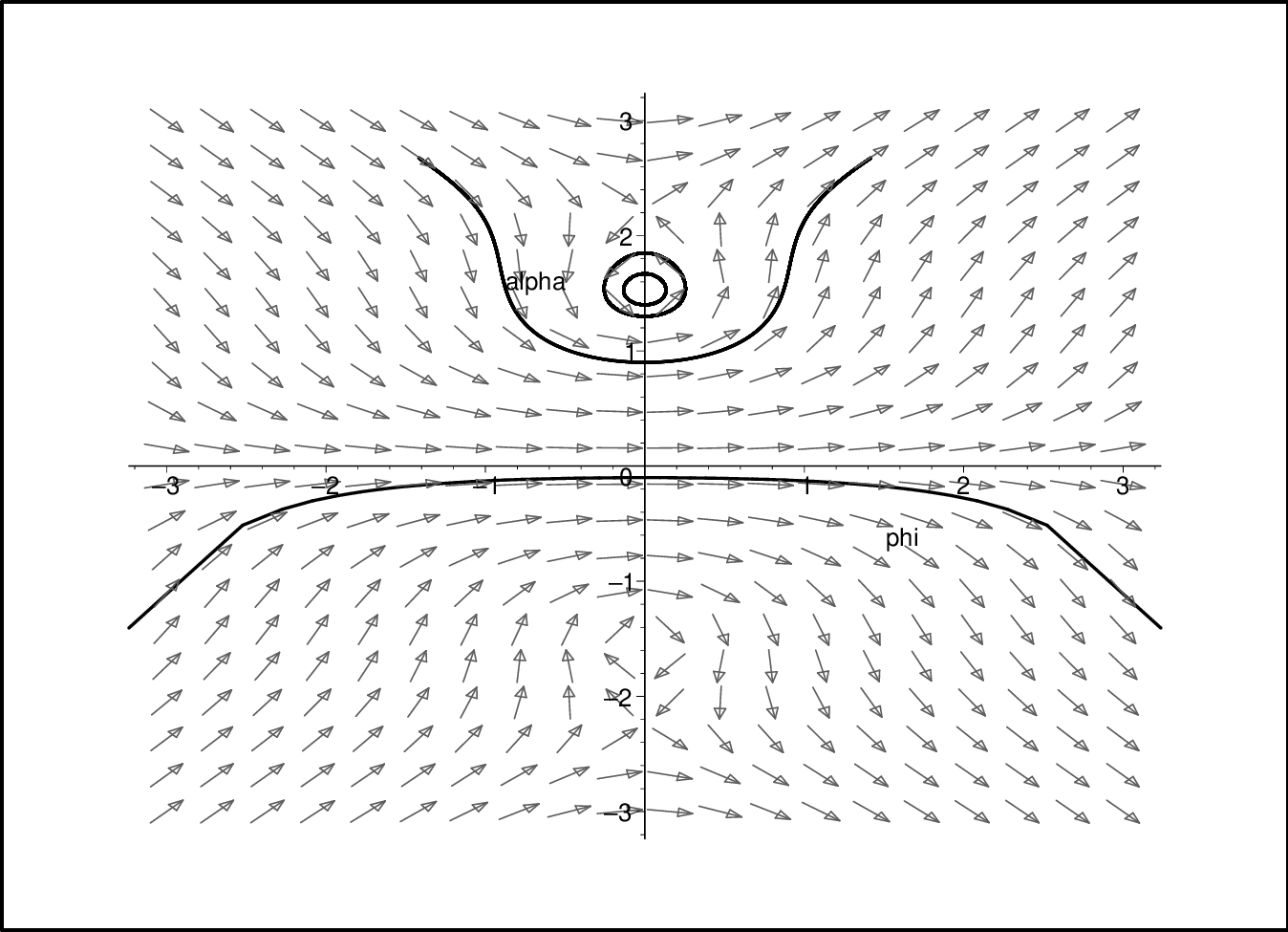}
\caption{The field plot shows the family of trajectories for the Bohm guidance equations (\ref{guia1}),(\ref{guia2}) associated to a wave functional with  the positive frequencies and  with such weigh of the negative frequencies that another cyclic universe emerges (trajectory passing through $\alpha=1.3 $, $\phi=0 $ ). $\sigma=1$ and $d=1$. }
\label{7}
\end{figure}

\begin{figure}[ht!]
\fbox{
\includegraphics[height=80mm,width=80mm]{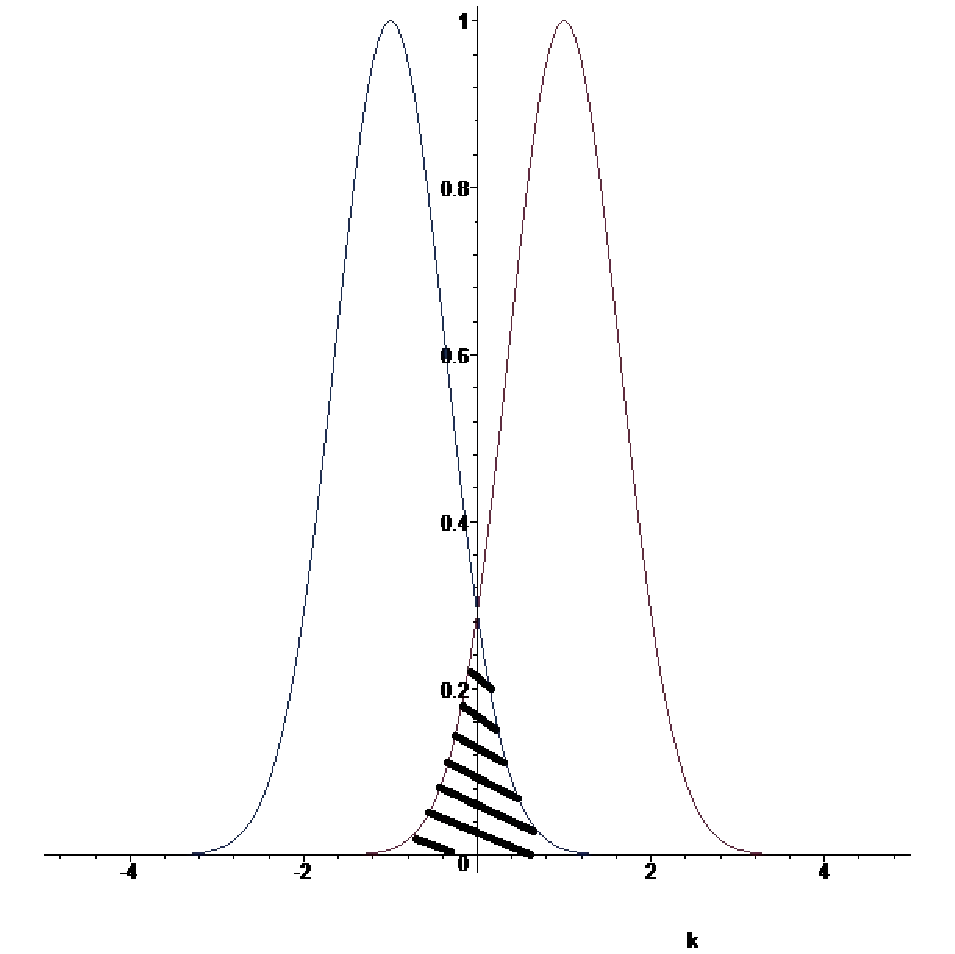}}
\caption{A given trajectory, potentially cyclic,  becomes effectively cyclic when the shaded area (which represent the contribution of the negative frequencies solutions) exceeds a threshold given by $T\approx  \sqrt{\pi}\sigma(1 - Erf(\frac{d}{\sigma})) \quad.$}
\label{umbral}
\end{figure}

We can ask whether there is a threshold of contribution of negative frequencies, above which a certain trajectory becomes cyclical, i.e. a threshold for the emergence of processes of creation-annihilation\footnote{I thank Prof. Nelson Pinto-Neto from CBPF, for asking along this line.}. As a partial answer we have found numerically that, for example, the trajectory whose initial conditions are $\alpha(0)=1.4 $, $\phi(0)=0 $ becomes cyclical when $\sigma \approx 0.9 $ for $d=1$ (Fig.\ref{6}). This can be characterized by the area enclosed under each Gaussian, between $ k = 0$ and $-\infty$ for the gaussian centered at $d$ and between $ k=0$ and $+\infty$ for the gaussian centered at $-d $, area that we call $T$ (Fig.\ref{umbral})\footnote{In other words, this is the sum of the areas of the tails along the  semi-axis which is opposite to the one containing the center of each gaussian.}: the threshold occurs for $T\approx \sqrt{\pi} \sigma(1 - Erf(\frac{d}{\sigma})) \approx 0.185$ ($\approx 5,8$ percent of the total area of the gaussians). There is another trajectory, with initial conditions $\alpha(0)=1.3 $, $\phi(0)=0 $, that becomes cyclical for $\sigma \approx 1 $ (Fig.\ref{7}), which means a threshold $T\approx  \sqrt{\pi}\sigma(1 - Erf(\frac{d}{\sigma})) \approx 0.279$ ($\approx 7,9$ percent of the total area of the gaussians). We see that the value of threshold $T$ depends strongly on initial conditions, i.e. it is different for  each of this type of trajectory.

At the point where a trajectory closes, we have $\frac{d\alpha}{d\phi}=0$ (the trajectory is instantaneously horizontal). This derivative is obtained by dividing equations (\ref{guia1}) and (\ref{guia2}), obtaining:

\begin{equation}\label{alfafi}
\frac{d\alpha}{d\phi}=\frac{\phi \sigma^2 \sin(\sqrt{2}d\alpha) +2\sqrt{2}d \sinh(\frac{\sigma^2\alpha\phi}{2})}
{2\sqrt{2}d \cosh(\frac{\sigma^2\alpha\phi}{2})+2\sqrt{2}d \cos(\sqrt{2}d\alpha)-\alpha\sigma^2 \sin(\sqrt{2}d\alpha)}
\end{equation}
In this way, knowing that at that point is $\phi = 0$, it is observed that the numerator and then derivative, is zero independently of the value of $\sigma$. This shows that there is no a generic value of $\sigma$ for which the trajectory closes, this will depend on the dynamics, as we will outline in the next section.
It is clear that of course not every trajectory becomes a cyclic universe by allowing  all the negative frequencies. It seems it would be possible to determine the set of trajectories that can become in cycling. Note that, for example, in the case of $\sigma = 0.9$ fixed, there may be more cyclical trajectories than indicated in Fig.\ref{6}, say that all trajectories interior at that. The same for fixed $\sigma = 1$.

\subsection{Qualitative physical explanation}

The transformation from open to closed trajectories can be qualitatively explained if we consider the energy balance of the system.

An equation that represents that  balance can be obtained from Eq.(\ref{hoqgp}) which is the quantum version of the Einstein-Hamilton-Jacobi equation (see \cite{peres}, \cite{wheeler}) and for our model is given by

\be
\label{hoqgp3}
- \biggl(\frac{\partial S(\phi, \alpha)}{\partial \alpha}\biggr)^2 + \biggl(\frac{\partial S(\phi, \alpha)}{\partial \phi}\biggr)^2
+ Q(\phi, \alpha) = 0 \quad,
\en

from which, using the guidance equations, we obtain

\be
\label{Tenergy}
0=\dot{\alpha}^{2}-\dot{\phi}^{2}
- \frac{Q}{\exp{6\alpha}}   \quad
\en

or equivalently, using $\alpha \equiv \log{a}$
\be
\label{Tenergya}
0=\biggl(\frac{\dot{a}}{a}\biggr)^{2}-\dot{\phi}^{2} 
- \frac{Q}{a^{6}}   \quad.
\en

Equation (\ref{Tenergy}), which is none other than the quantum version of the Friedmann equation, can be interpreted as representing a system with the total energy (given by the LHS, which is constant and in this case is zero for having assumed a flat geometry) equal to the sum of the "kinetic energy", represented by the first term of the RHS, plus the "effective potential energy" given by the other terms of the RHS. This resembles to a particle moving under the action of a classical potential, although not exactly equivalent, since the "`effective potential energy" ($=-\dot{\phi}^{2}
- \frac{Q}{\exp{6\alpha}} $) includes the quantum potential, which is strongly non-local and non-linear.


We know that the singular points, i.e centers and nodes appear along the axis $\phi = 0$. Then, in order to have an qualitative idea of what is happening, we can consider $\phi = constant$ (a small value but not zero because in this case the equation reduces to $0=0$)   so we have a system dependent on only one coordinate, namely $\alpha$.  

\begin{figure}[!ht]
\includegraphics[height=80mm,width=80mm,angle=270]{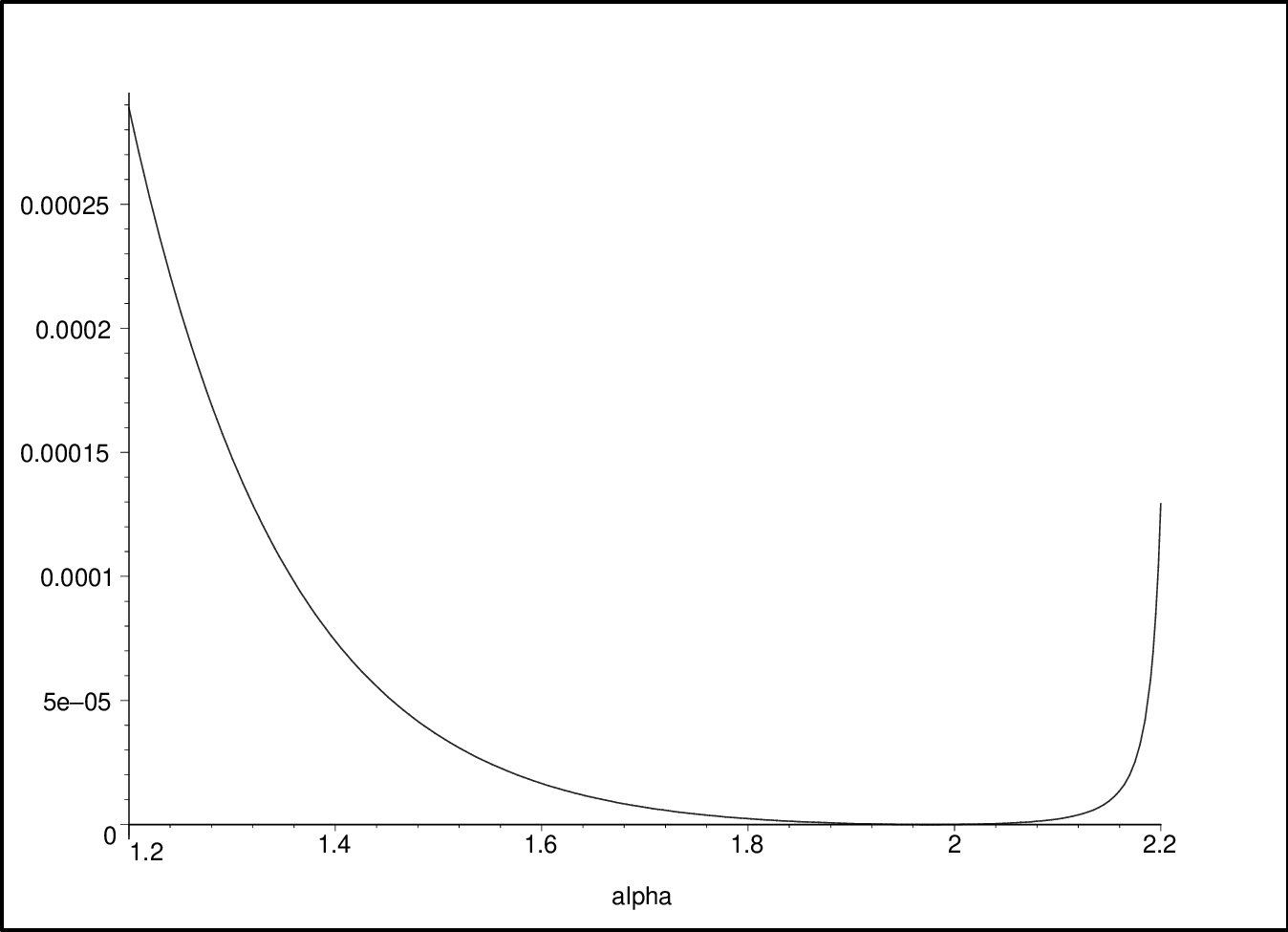}
\caption{The minimum of the efective potential for $\sigma=0.5$ is at $\alpha=1.9768$. The pole is fixed at $\alpha=2.22$. }
\label{Vq5}
\end{figure}

\begin{figure}[!ht]
\includegraphics[height=80mm,width=80mm,angle=270]{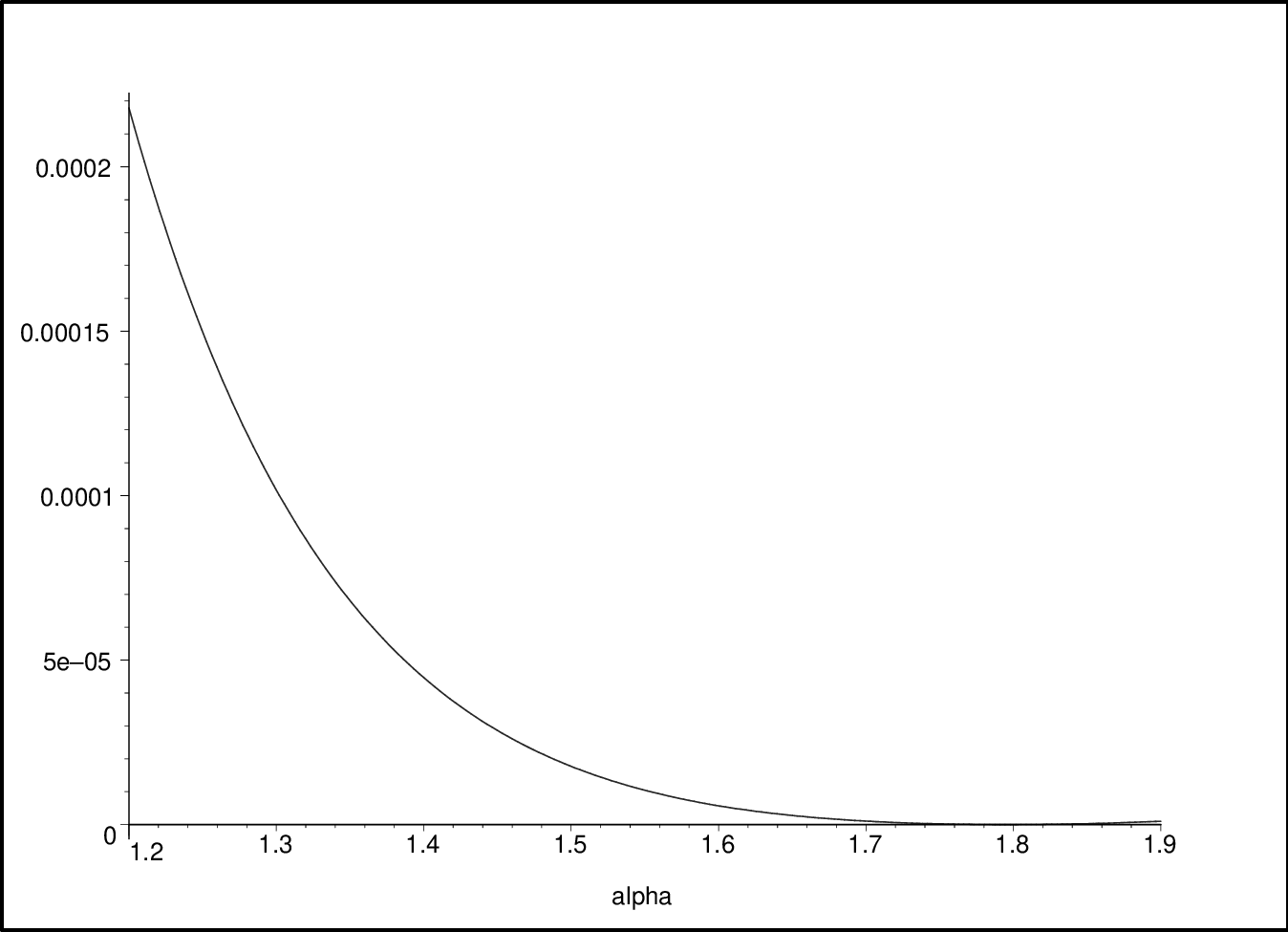}
\caption{The minimum of the efective potential for $\sigma=0.7$ is now at $\alpha=1.795$, i.e, it moved away from the node which remains fixed at $\alpha=2.22$.}
\label{Vq7}
\end{figure}

\begin{figure}[!ht]
\includegraphics[height=80mm,width=80mm,angle=270]{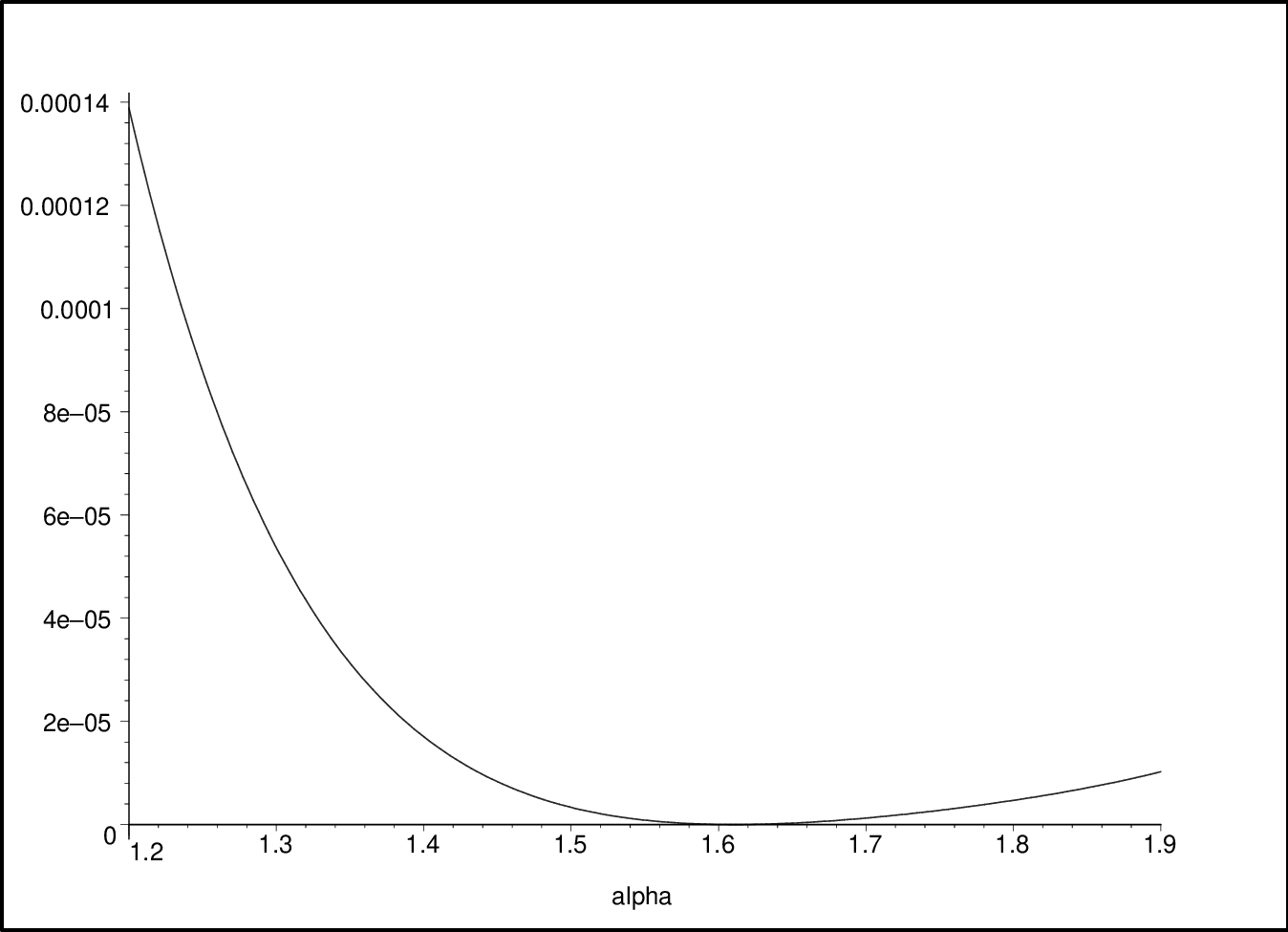}
\caption{The minimum of the efective potential for $\sigma=0.9$ has shifted to $\alpha=1.6104$. The trajectory with initial conditions $\alpha(0)=1.4$, $\phi(0)=0$ "has found a place" to close.}
\label{Vq9}
\end{figure}

\begin{figure}[!ht]
\includegraphics[height=80mm,width=80mm,angle=270]{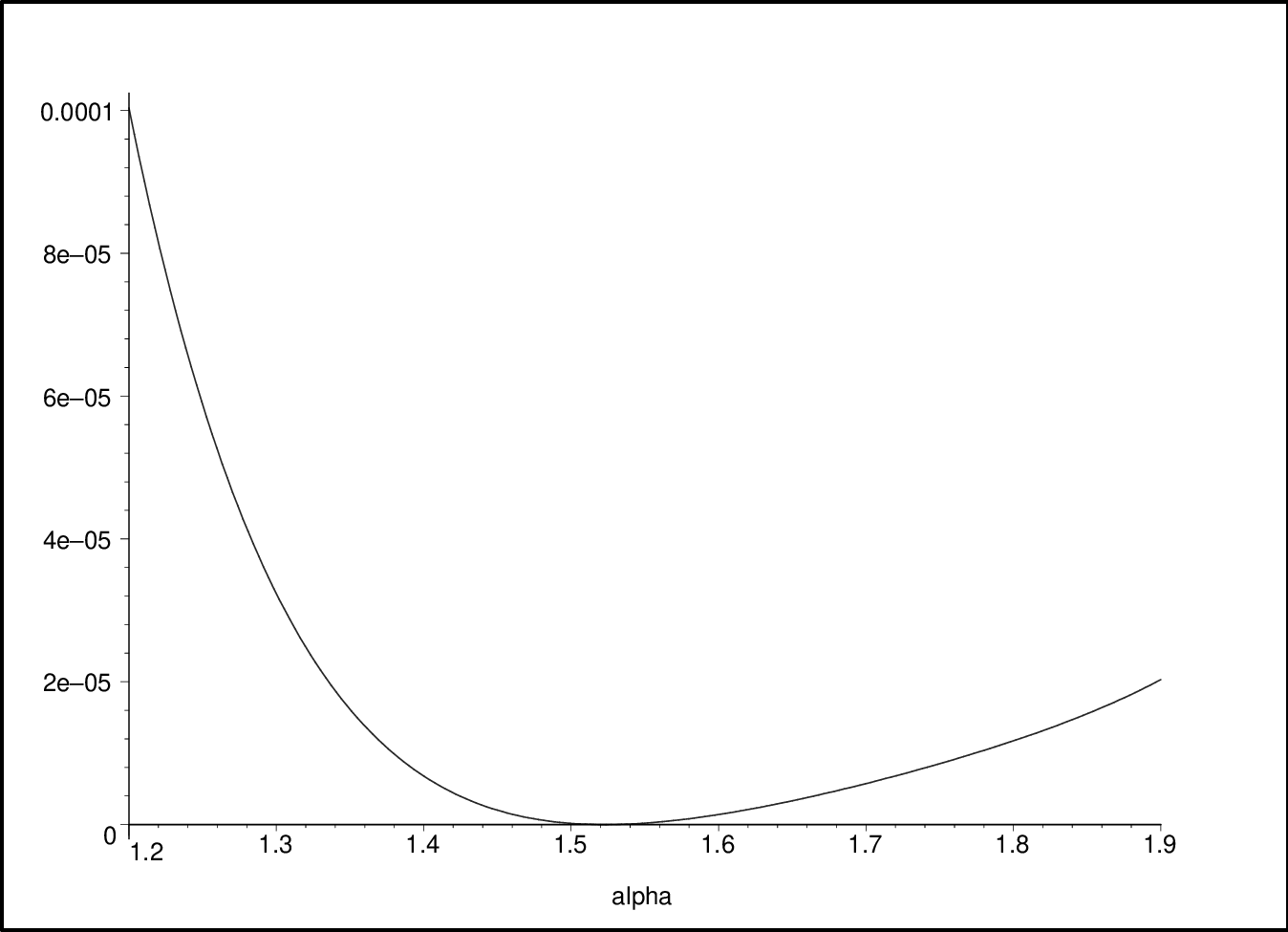}
\caption{The minimum of the efective potential for $\sigma=1$ has shifted to $\alpha=1.522$ and another trajectory becomes closed (the one with initial conditions $\alpha(0)=1.3$, $\phi(0)=0$.}
\label{Vq1}
\end{figure}

 Then the local minimums of  $V_{q}\equiv -\frac{Q}{\exp{6\alpha}}$ will define, as in a dynamical system, the centers, and the poles will define  the unreachable regions, that is, the nodes. In figures \ref{Vq5}, \ref{Vq7},\ref{Vq9}, and \ref{Vq1} are represented a cut of the effective potential as a function of $\alpha$.

We see that as parameter $\sigma$ increases, the position of the minimum (which, as we said, defines the centers) is moving away from the node (which, as we know, remains fixed) so that there are open trajectories that "acquire space" to close (the node prevented it). This could be explained as in a dynamic system through the energy balance but now we have the quantum potential, which is strongly non-linear in the considered region and makes it difficult to obtain an analytical result.

\section{Conclusion}
We have considered the procedure of discarding negative frequencies solutions, usual in quantum cosmology and which is made invoking a type of  ``super-selection''. 
The discarding of negative frequency solutions  in a QFT brings about the absence of antiparticles which, after all, means the violation of 4-inversion symmetry $(x \rightarrow -x, t \rightarrow-t)$ which is a (improper) Lorentz transformation. As an heuristic discussion suppose you have a theory of quantum gravity  which lacks the  negative frequency solutions. Taking some limit in this theory in order to obtain the weak (or null)  gravitational  regime,  the result is a theory that does not respect Lorentz symmetry and does not have place for antiparticles. That is, a relativistic QFT is not obtained, as it should be. 

For the case of a quantum cosmology model we have shown that if we ignore the negative frequency solutions,  the rich processes of creation/annihilation of universes at the Planck scale, are lost. 
In fact, we were able to obtain the bohmian trajectories given by solutions of the Wheeler-DeWitt equation for a simple model. We consider initially a positive frequency solution and we have studied numerically the behavior of trajectories while
were including the negative frequencies. We have shown that when the negative frequencies are considered on an equal footing than positive frequencies, as the general solution of any Klein-Gordon type equation requires, new processes, prior absent, appear: cyclic universes of Planckian size, which can be interpreted as processes of creation-annihilation of universes that exist for a very short duration. This is a natural feature of any quantum  relativistic fields theory. In this way our results  have led us to believe  that this super-selection rule can not exist.

We verified numerically that, for a given trajectory,  there is a threshold of negative frequencies solutions, above which cyclic universes are obtained, i.e. processes of creation-annihilation. We see that this depends strongly on initial conditions, i.e. it is different for  each of this type of trajectory.  Moreover, it is clear that not every trajectory becomes a cyclic universe by adding  the negative frequencies. However it could be possible to determine the set for which this is possible. This could be  a future  topic of research. 
Another important point for a new investigation would be to analyze the more general case of a potential $V (\alpha,\phi)$,
in which the null coordinates in WDW equation are not completely separated.

\section{Acknowledgments}

I would like to thank CNEN and CBPF from MCTI Brasil for their support. I also wish to thank Professor Sebasti\~ao Alves Dias from CBPF for helpful comments and clarifications on quantum field theory.

\end{document}